\documentclass[
  prx,
  tightenlines,
  aps,
  nofootinbib,
  superscriptaddress,
  longbibliography,
  twocolumn,
  10pt
]{revtex4-2}

\usepackage{amsmath,amssymb,amsthm,bm,array,graphicx}
\usepackage[caption=false]{subfig}
\usepackage[export]{adjustbox}
\usepackage[pdftex,pagebackref=false]{hyperref}
\usepackage{xcolor}
\usepackage{tikz}
\usetikzlibrary{arrows.meta,calc}

\hypersetup{
  colorlinks=true,
  linkcolor=blue,
  citecolor=blue,
  urlcolor=blue,
  unicode=true
}
\usepackage{physics}
\usepackage{cleveref}
\usepackage{booktabs}
\usepackage{threeparttable}   
\newcommand{\ii}{\mathrm{i}}

\newcommand{\Repart}{\operatorname{Re}}
\newcommand{\Impart}{\operatorname{Im}}

\DeclareMathOperator{\triu}{triu}
\DeclareMathOperator{\rev}{rev}
\DeclareMathOperator{\Var}{Var}

\newcommand{\btheta}{\boldsymbol{\theta}}

\newcommand{\ba}{\boldsymbol{a}}
\newcommand{\bx}{\boldsymbol{x}}
\newcommand{\by}{\boldsymbol{y}}

\newcommand{\bF}{\boldsymbol{F}}

\newcommand{\bv}{\boldsymbol{v}}
\newcommand{\bzero}{\boldsymbol{0}}

\newcommand{\mcE}{\mathcal{E}}

\newcommand{\mcK}{\mathcal{K}}

\newcommand{\ppsi}{\psi_{\btheta}}

\theoremstyle{definition}

\begin{document}
\title{Numerical simulation of D-Wave's quantum advantage experiment with time-dependent variational Monte Carlo}
\author{Roeland Wiersema}
\affiliation{Center for Computational Quantum Physics, Flatiron Institute, 162 Fifth Avenue, New York, NY 10010, USA}
\date{\today}

\begin{abstract}
Programmable quantum annealers can realize real-time dynamics of frustrated transverse-field Ising models on large, nontrivial graphs. Recent work by King et al. argued that the classical simulation of such experiments would require exponential computational resources for classical methods such as tensor networks and neural quantum states. 
Here, we numerically simulate the D-Wave spin-glass annealing protocol with time-dependent variational Monte Carlo (t-VMC) using a correlator state tailored to spin-glass dynamics. 
For the two-dimensional cylinder, three-dimensional dimer, diamond, and biclique instances considered, at annealing times of 7 and 20 ns, we show that systematically increasing the variational ansatz size enables t-VMC to approximate the final two-spin correlation errors of the quantum processing unit (QPU). We also perform an accurate large-scale simulation of a challenging biclique instance for which no other variational method is known to work, and we find close agreement with the quantum annealer.
Through ablation studies, we identify poor Markov-chain mixing, high-variance local-energy estimators, and stochastic Runge--Kutta error estimates as the principal numerical failure modes. We address these numerical issues by using parallel tempering, blurred sampling and an importance-weighted differential equation solver, thereby clarifying the numerical requirements for stable, large-scale t-VMC simulations.
Our results extend the frontier of classical simulation while providing a realistic assessment of the computational costs of simulating quantum dynamics at this scale.
\end{abstract}

\maketitle

\section{Introduction}

Accurately simulating the nonequilibrium dynamics of interacting quantum
many-body systems remains a central challenge for classical computational
methods. Programmable quantum annealers can realize dynamics in disordered transverse-field Ising models at system sizes inaccessible to exact state-vector simulation~\cite{Harris2018PhaseTransitions,Bando2020ProbingUniversality,
King2022CoherentAnnealing,King2023QuantumCriticalDynamics,
Manovitz2025QuantumCoarsening,Zhang2025NearCriticalKZ,
Fang2025ProbingCriticalPhenomena}. 
In a prototypical annealing protocol, the system is initialized in a
field-polarized state and driven toward a frustrated Ising spin-glass
Hamiltonian.
During this evolution, the system traverses a quantum-critical
regime in which correlations spread over increasingly large length
scales. 
The resulting dynamics present a stringent
test for classical simulation: the exponential growth of the Hilbert space
is compounded by quenched disorder, frustration, critical correlations, and
the nontrivial interaction geometry of programmable hardware.

Using D-Wave's Advantage2 processor,
Ref.~\cite{King2025BeyondClassical} reported samples in close agreement
with converged classical calculations on classically accessible
spin-glass instances. 
At larger sizes, the experiment reproduced
expected quantum-critical scaling on 2D cylinder, 3D dimer, diamond, and
biclique topologies. 
Comparisons with matrix-product states (MPS),
projected entangled-pair states (PEPS), and neural quantum states (NQS)
led the authors to argue that the classical methods tested could not
reproduce the same observables at comparable cost for the largest
systems.

As was the case for previous transverse-field Ising quantum simulation experiments~\cite{kim2023evidence, Tindall2024EfficientTensorNetwork, Begusic2024FastConverged, Patra2024LargestProcessors, Park2025InfluenceFunctionalBP}, 
subsequent classical work has substantially refined this frontier.
Geometry-matched tensor networks in two and three dimensions have
reproduced the disordered annealing dynamics on 2D cylinder, 3D dimer, and
diamond lattices, including their universal critical behavior~\cite{Tindall2026TensorNetworks}. 
However, it was shown that these methods can struggle on bimodal disorder and are not clearly extendable to the highly connected biclique lattices explored in the original experiment~\cite{nocera2025evaluating}.
For biclique spin glasses, a discrete
truncated-Wigner approximation has reproduced instance-to-instance
fluctuations of the Edwards--Anderson order parameter and the
corresponding Binder-cumulant scaling at low computational cost~\cite{Sels2026TruncatedWigner}. 

Variational Monte Carlo (VMC) offers a complementary family of approaches, including time-dependent VMC (t-VMC)~\cite{Ido2015tvmc, Carleo2017Neural,Schmitt2020QuantumDynamics,Schmitt2022quantum, Nys2024ab, Salioni2026adaptive, medvidovic2025adiabatic, Vu2025optimizing, naik2026real}, projected VMC~\cite{Medvidovic2021classical,Gutierrez2022RealTime,Donatella2023dynamics,Gravina2025NeuralProjected,Sinibaldi2023Unbiasing}, and more recently explicitly parameterized t-NQS methods~\cite{van2025many, Sinibaldi2026NeuralGalerkin}.
Using these methods, variational simulations of quantum dynamics have been explored across a broad range of ans\"atze and physical settings~\cite{schmitt2025simulating}, often requiring protocols carefully tailored to the physics under study.
These developments motivate a systematic study of the algorithmic ingredients required for VMC to reproduce quantum-annealing dynamics in frustrated spin systems. 
In Ref.~\cite{King2025BeyondClassical}, a t-VMC calculation based on a complex restricted Boltzmann machine reproduced the correlations of the 2D cylinder instances; however, this ansatz was found to be challenging to optimize and regularize at scale. 
More recently, Ref.~\cite{Mauron2025Challenging} reported a successful simulation of the $7$ ns diamond instances.

In this work we present a comprehensive numerical study for each topology considered in the original experiment of Ref.~\cite{King2025BeyondClassical}. We obtain an error on par with the QPU for the $t_a=7$ ns and $t_a=20$ ns instances across all system sizes for which converged MPS ground truth data are available. Additionally, we perform a large-scale simulation of a $N=72$ biclique instance. Here, no other variational method is known to produce a correct state but we find close agreement with the QPU result. We also simulate a $N=128$ diamond instance in the challenging $20$ ns regime to demonstrate the scalability of our approach.

Three numerical ingredients are central to these calculations. First, we introduce a structured correlator ansatz whose expressivity can be increased systematically through a single channel-rank hyperparameter $R$. Second, we use blurred sampling to control the high-variance and support-mismatch pathologies that arise in local-energy estimators~\cite{Wan2026BlurredSampling}. Finally, we reuse and reweight Monte Carlo samples within each adaptive Runge--Kutta step required for solving the TDVP equation, thereby preventing sampling noise from dominating the embedded error estimate.

\begin{figure*}[htb!]
    \centering
    \input{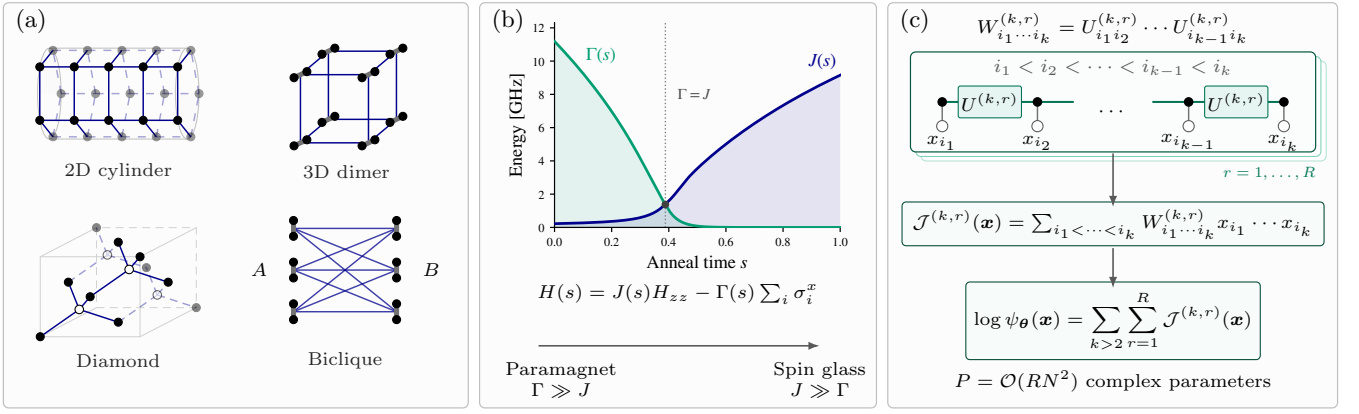}
    \caption{
    Overview of the quantum-annealing simulation.
    (a)~Representative 2D cylinder, 3D dimer, diamond, and biclique problem graphs.
    (b)~Annealing schedule interpolating between the transverse-field paramagnet and the spin-glass Hamiltonian.
    (c)~Path-factorized $k$-body correlator ansatz with channel rank $R$ and $\mathcal{O}(RN^2)$ complex parameters. We omit the global phase and odd terms due to symmetry considerations.
    }
    \label{fig:overview}
\end{figure*}

\section{Quantum Annealing of Spin Glasses}
\label{sec:background}

The quantum-annealing protocol of Ref.~\cite{King2025BeyondClassical} evolves the transverse-field Ising model (TFIM) on $N$ spins,
\begin{align}
  H(s)
  &= J(s)\sum_{(i,j)\in\mcE}w_{ij}\,\sigma_i^z\sigma_j^z-\Gamma(s)\sum_{i=1}^{N}\sigma_i^x,
  \label{eq:H}
\end{align}
Here $s\in[0,1]$ is the dimensionless anneal parameter, $\mcE$ is the edge set of the problem graph, and $\boldsymbol{w}=\{w_{ij}\}_{(i,j)\in\mcE}$ specifies the spin-glass instance. The schedule functions $J(s)$ and $\Gamma(s)$ follow the experimental annealing schedule (see Fig.~\ref{fig:overview}(b)). Physical time is related to the anneal parameter by $t=t_a s$, where $t_a$ is the total annealing time.

For sufficiently large $t_a$, the evolution approaches the adiabatic limit and ends near the ground-state manifold of the Ising part of the Hamiltonian. At a finite rate, however, critical slowing down prevents the state from following the instantaneous ground state through the quantum-critical region.
The Kibble--Zurek mechanism relates the resulting freeze-out length and the density of excitations or defects to the ramp time through the equilibrium critical exponents~\cite{Zurek1985Cosmological,Polkovnikov2005Adiabatic,delCampoZurek2014,Dziarmaga2006Random}.
For the present schedule, the crossing of the nominal transverse and Ising energy scales, $\Gamma(s)\simeq J(s)$, provides a useful landmark for the onset of strongly correlated dynamics. We consider the 2D cylinder, 3D dimer, diamond, and biclique topologies from Ref.~\cite{King2025BeyondClassical}, which are shown in Fig.~\ref{fig:overview}(a) and summarized in App.~\ref{app:topologies}.

\section{Method}
\label{sec:method}

We simulate the nonequilibrium dynamics with time-dependent variational Monte Carlo. In the computational basis $\{\ket{\bx}\}$, with $\bx\in\{-1,+1\}^{N}$, the variational state and its Born distribution are
\begin{equation*}
  \ket{\psi_{\btheta}}
  = \sum_{\bx}\psi_{\btheta}(\bx)\ket{\bx},
  \qquad
  p_{\btheta}(\bx)
  = \frac{\abs{\psi_{\btheta}(\bx)}^2}
  {\sum_{\by}\abs{\psi_{\btheta}(\by)}^2},
\end{equation*}
where $\btheta\in\mathbb{C}^{P}$ contains the $P$ complex variational parameters. We define the logarithmic derivatives $O_{\mu}(\bx)= \partial_{\theta_{\mu}}\log\psi_{\btheta}(\bx)$
and the local energy
\begin{equation}
  E_{\mathrm{loc}}(\bx;s)
  = \frac{\bra{\bx}H(s)\ket{\psi_{\btheta}}}
  {\psi_{\btheta}(\bx)}.
  \label{eq:local_energy}
\end{equation}
The covariance forms of the quantum geometric tensor (QGT) and the force vector are
\begin{align}
  S_{\mu\nu}(\btheta)
  &=
  \left\langle
  \left(O_{\mu}-\langle O_{\mu}\rangle\right)^{*}
  \left(O_{\nu}-\langle O_{\nu}\rangle\right)
  \right\rangle_{p_{\btheta}}\notag\\
  F_{\mu}(s, \btheta)
  &=
  \left\langle
  \left(O_{\mu}-\langle O_{\mu}\rangle\right)^{*}
  E_{\mathrm{loc}}(\bx;s)
  \right\rangle_{p_{\btheta}}.
  \label{eq:force}
\end{align}
Projecting the Schr\"odinger equation onto the tangent space of the variational manifold gives the t-VMC equation of motion~\cite{McLachlan1964Variational,Carleo2012Localization, Hackl2020geometry},
\begin{equation}
  S(\btheta)\dot{\btheta}=-\ii\,\bF(s,\btheta).
  \label{eq:tdvp-main}
\end{equation}
The initial state at $s=0$ is the variational approximation to the ground state of $H(0)$, which is close to the transverse-field paramagnet. We obtain the initial parameters by stochastic reconfiguration~\cite{Sorella1998green} and stop when the V-score is below $10^{-4}$~\cite{Wu2024variational}.

After the anneal, the primary observables of interest are the two-spin Ising correlations
\begin{equation*}
  C_{ij}
  =
  \frac{
    \bra{\psi_{\btheta}}\sigma_i^z\sigma_j^z
    \ket{\psi_{\btheta}}
  }{
    \braket{\psi_{\btheta}}{\psi_{\btheta}}
  }.
\end{equation*}
We define the relative correlation error with respect to the reference data as
\begin{equation}
  \epsilon_c
  =
  \frac{
    \sqrt{
      \sum_{i>j}^N
      \abs{C_{ij}^{\mathrm{VMC}} - C_{ij}^{\mathrm{ref}}}^2
    }
  }{
    \sqrt{
      \sum_{i>j}^N
      \abs{C_{ij}^{\mathrm{ref}}}^2}
  }.
  \label{eq:corr-error}
\end{equation}
The reference data for the instances studied here are the converged MPS calculations reported in Ref.~\cite{King2025BeyondClassical}. We use $\epsilon_c<5\%$ as an operational accuracy target motivated by the scale of the QPU--MPS discrepancies reported there.

Achieving this accuracy at practical cost requires coordinated choices of variational ansatz (\Cref{sec:ansatz}), sampler (\Cref{sec:sampler}), and TDVP solver (\Cref{sec:solve}), which we explore in detail in the subsequent sections. Additionally, we briefly summarize the experiments that did not lead to improvements in the results in App.~\ref{app:failed}. Numerical simulations were performed with NetKet~\cite{netket2:2019, netket3:2022}, a VMC framework built on JAX~\cite{jax2018github} and Flax~\cite{flax2020github}.

All code, data, and simulation checkpoints are publicly available~\cite{code}. The hyperparameters and wall times for each experiment can be found in App.~\ref{app:hyperparameters}.

\subsection{Variational ansatz}
\label{sec:ansatz}

A successful t-VMC calculation requires a variational family that is expressive enough to represent the evolving state while retaining a tangent-space geometry that can be regularized reliably. We use a specific set of correlator states,
\begin{equation}
  \log\psi_{\btheta}(\bx)
  = c+\sum_i a_i x_i
  + \sum_{k\in\mcK}\sum_{r=1}^{R_k}\mathcal{J}^{(k,r)}(\bx),
  \label{eq:ansatz}
\end{equation}
with
\begin{equation*}
  \mathcal{J}^{(k,r)}(\bx)
  =
  \sum_{1\leq i_1<\cdots<i_k\leq N}
  W^{(k,r)}_{i_1\cdots i_k}
  x_{i_1}\cdots x_{i_k}.
\end{equation*}
Related variational families have been studied in both VMC and tensor-network settings~\cite{Changlani2009CPS,Marti2010complete,Mezzacapo2009ground,Mauron2025Challenging}. Here the $k$-body coefficient tensor is constrained by the path factorization
\begin{equation}
  W^{(k,r)}_{i_1\cdots i_k}
  = U^{(k,r)}_{i_1 i_{2}}\cdots U^{(k,r)}_{i_{k-1} i_{k}},
  \label{eq:path-factor}
\end{equation}
where each $U^{(k,r)}\in\mathbb{C}^{N\times N}$ is strictly upper triangular (see App.~\ref{app:filltri} for a fast and differentiable implementation). A single channel therefore represents all ordered $k$-spin monomials, while tying their coefficients through paths along the ordered tuple $i_1<\cdots<i_k$. We refer to $R_k$ as the channel rank for body order $k>3$, similar in spirit to the rank of a Canonical Polyadic decomposition~\cite{Sorber2013}.

For fixed $k$ and $R_k$, the factorization reduces the number of independent complex parameters from $\mathcal{O}(N^k)$ to $\mathcal{O}(R_kN^2)$. 
Throughout this work, we use a common channel rank $R_k=R$ and body orders $\mcK=\{2,4\}$. The $\mathbb{Z}_2$ symmetry of Eq.~\eqref{eq:H} allows us to set $\ba=\bzero \in \mathbb{C}^N$. We fix $c=0$ because the overall normalization and global phase are physically irrelevant.

\subsection{Sampling the variational state}
\label{sec:sampler}

At large $s$, the target distribution $p_{\btheta}(\bx)$ becomes glassy and multimodal. Single-spin Metropolis proposals then decorrelate slowly, and independent chains can become trapped in different modes. This slowdown is expected for rugged spin-glass distributions, where
local updates can remain trapped behind free-energy barriers~\cite{Binder1986spinglass}. While autoregressive variational states can circumvent sampling difficulties~\cite{Hibat2021variational,Inack2022neural,Mcnaughton2020boosting}, the ansatz of Eq.~\eqref{eq:ansatz} is unnormalized and must be sampled with Markov chain Monte Carlo. To prevent the critical slowing down of sampling in the glassy regime, we use parallel tempering, also known as replica exchange Monte Carlo, to sample the multimodal distributions that appear late in the anneal~\cite{Marinari1992simulated, Hukushima1996ExchangeMC, Earl2005parallel}. For a fixed variational state, define $N_T$ tempered replicas with inverse-temperature ladder $(\beta_0,\ldots,\beta_{N_T-1})
$ with $1=\beta_0>\beta_1>\cdots>\beta_{N_T-1}\ge 0$. This defines the target distributions
\begin{equation*}
  \pi_m(\bx)
  \propto p_{\btheta}(\bx)^{\beta_m}
  = \abs{\psi_{\btheta}(\bx)}^{2\beta_m}.
\end{equation*}
The $\beta_0=1$ replica samples the physical distribution, while replicas at smaller $\beta_m$ explore a flatter landscape and can cross free-energy barriers more readily.

Each replica performs local Metropolis sweeps targeting its own $\pi_m$. After a fixed number of sweeps, neighboring replicas $m$ and $m+1$ propose to exchange configurations $\bx_m$ and $\bx_{m+1}$. We define the log-density difference $\Delta\ell_m
  = 2\Repart\left[
    \log\ppsi(\bx_{m+1})-
    \log\ppsi(\bx_m)
  \right]$.
Detailed balance for the joint distribution $\prod_m\pi_m(\bx_m)$ then gives the acceptance ratio between replica states as
\begin{equation*}
  \alpha^{m, m+1}_{\mathrm{swap}}
  = \min\left\{1,
    \exp\left[(\beta_m-\beta_{m+1})\Delta\ell_m\right]
  \right\}.
\end{equation*}
Only samples from the physical replica at $\beta_0=1$ enter observables and TDVP estimators. Nevertheless, replica exchange allows a configuration to diffuse up and down the temperature ladder, crossing energy barriers and recovering physical modes.

\begin{figure*}[htb!]
    \centering
    \includegraphics[width=\linewidth]{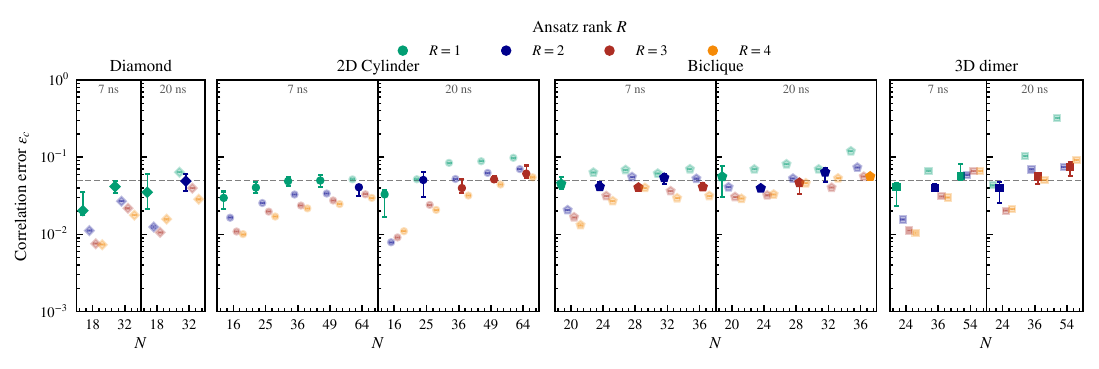}
    \caption{Correlation error of the t-VMC ansatz at the end of the anneal. Final-time ($s=1$) ZZ-correlation error $\epsilon_c$ relative to the MPS reference, for every simulated topology. Within each topology, we show the two anneal times $t_a=7$ ns and $t_a=20$ ns. 
    The markers are colored by ansatz rank $R\in\{1,2,3,4\}$ and horizontally offset by rank. The correlation error $\epsilon_c$ is estimated using $2^{21}=2097152$ samples. To assess stability across disorder realizations, we repeat the optimization for five instances at selected ranks that reach approximately $5\%$ error (dashed line); error bars indicate the minimum and maximum, and the marker indicates the median.
    }
    \label{fig:final}
\end{figure*}

In addition to mixing problems of the MCMC sampler, the estimators entering the TDVP equation suffer from statistical pathologies as the state traverses the critical point. In particular, the force estimator in Eq.~\eqref{eq:force}, contains amplitude ratios between Hamiltonian-connected configurations. Near nodes or in regions of very small probability, these ratios can have extremely large variance. Additionally, exact zeros can also produce support-mismatch bias in standard estimators~\cite{Sinibaldi2023Unbiasing,Krinitsin2026TVMCWithoutBias,Wan2026BlurredSampling,Chen2026SupportMismatchRotation}. We address this pathology with a blurred proposal distribution that was introduced in Ref.~\cite{Wan2026BlurredSampling}. 

Starting from $\bx\sim p_{\btheta}$, we draw a blurred configuration $\by$ from
\begin{equation*}
  K(\by|\bx)
  = (1-q)\delta_{\by,\bx}+qK_{\mathrm{off}}(\by|\bx),
  \qquad
  \sum_{\by}K(\by|\bx)=1,
\end{equation*}
where $K_{\mathrm{off}}$ is the uniform single-spin-flip kernel induced by the off-diagonal term in of the annealing Hamiltonian in Eq.~\eqref{eq:H}. The marginal blurred density is
\begin{equation*}
  r_{\btheta}(\by)
  = \sum_{\bx}p_{\btheta}(\bx)K(\by|\bx).
\end{equation*}
Expectations of an observable $\mathcal{A}(\bx)$ with respect to $p_{\btheta}$ are then estimated using self-normalized importance sampling,
\begin{equation}
  \langle \mathcal{A}\rangle_{p_{\btheta}}
  \approx
  \frac{
    \sum_{n=1}^{N_s}w_n \mathcal{A}(\by_n)
  }{
    \sum_{n=1}^{N_s}w_n
  },
  \qquad
  w_n
  = \frac{p_{\btheta}(\by_n)}
  {r_{\btheta}(\by_n)}.
  \label{eq:blur-reweight}
\end{equation}

\subsection{Cached Runge--Kutta solver}
\label{sec:solve}

The TDVP update requires the solution of an ill-conditioned linear system. Small eigenvalues of $S(\btheta)$ correspond to tangent-space directions that are redundant, weakly represented by the current state, or unresolved by the finite Monte Carlo sample. We therefore use the smooth regularized pseudoinverse $S^{+}_{\mathrm{reg}}(\btheta)$ according to the regularization scheme of Ref.~\cite{Schmitt2020QuantumDynamics}. Solving the linear system of~\eqref{eq:tdvp-main} then gives the differential equation,
\begin{equation}
  \dot\btheta
  = -\ii\,S^{+}_{\mathrm{reg}}(\btheta)\bF(s,\btheta) \equiv \boldsymbol{f}(s, \btheta),
  \label{eq:regularized-tdvp}
\end{equation}
which is a time-dependent ordinary differential equation (ODE). To solve this ODE, we use a $p$-th order Runge--Kutta (RK) method. In general, such an approach is described by the coefficients of the RK matrix $A$ and the weights and nodes $\boldsymbol{b}, \,\boldsymbol{c}$, respectively ~\cite{Butcher2016}. Given these coefficients, an ODE update with time step $\Delta s$ can be computed as
\begin{align*}
  \bar \btheta^{(p)}
  &=
  \btheta + \Delta s\sum_{i=1}^{n_{\mathrm{st}}}b_i^{(p)}\boldsymbol{k}_i, \qquad \boldsymbol{k}_i = \boldsymbol{f}\big(s +\Delta s\,c_i ,\btheta_i\big)
\end{align*}
with the parameters at RK stage $i$ given by $\btheta_i= \btheta + \Delta s\sum_{j<i}^{n_{\mathrm{st}}}a_{ij}\boldsymbol{k}_j$.
We then evaluate the right-hand side at $n_{\mathrm{st}}$ tableau stages and construct two updates of different orders. The difference between the two updates, $\delta\bar\btheta^{(p)}=\bar\btheta^{(p)}-\bar\btheta^{(p-1)}$, can be used to estimate the local truncation error. In particular, if we define the QGT-induced norm as $\|\bv\|_{S}:=\sqrt{\bv^\dagger S\bv}$, we can calculate the dimensionless error quantity
$ \eta=\|\delta\bar\btheta^{(p)}\|_{S}/\|\btheta\|_{S}$,
which is then used by an adaptive controller to adjust the step size $\Delta s$ when $\eta\leq\texttt{rtol}$.

For the stochastic TDVP right-hand side
$\boldsymbol{f}(s,\btheta)$, drawing an independent Monte Carlo
sample at every RK stage causes the embedded difference
$\delta\bar\btheta^{(p)}$ to be dominated by resampling noise rather than
by deterministic integration error. The adaptive controller can
then become pinned near its minimum allowed step size, substantially
increasing the integration cost. 
\begin{figure*}[htb!]
  \centering
  \includegraphics[width=\linewidth]{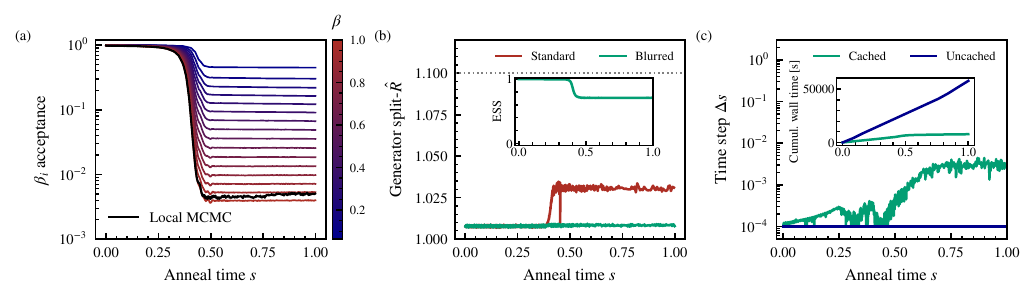}
  \caption{Ablation study of methodological components of t-VMC simulation.
  (a)~Local MCMC dynamics versus replica exchange Monte Carlo. At low temperatures, acceptance rates freeze within each chain, while at high temperatures, acceptance rates remain reasonable. 
  The adjacent-replica exchange rate fluctuates between 0.239 and 0.288, enabling configurations to diffuse along the temperature ladder and reach the physical $\beta_0$ replica. The dashed line shows the freezing of the local MCMC dynamics. The final error is $\epsilon_c = 0.4096 \pm 0.0027$ for local sampling, whereas for replica exchange we recover the $<5\%$ error $\epsilon_c = 0.0331 \pm 0.0003$.
  (b)~Standard sampling versus blurred sampling. With the standard estimator, the split-$\widehat{R}$ for the instantaneous energy of $H(s)$ deviates from unity near the crossover region, whereas blurred sampling remains well converged. Inset: effective sample size of the blurred estimator, calculated as $\mathrm{ESS}=(\sum_n w_n)^2/\sum_n w_n^2$. We find that the standard sampler results in an error of $\epsilon_c=0.2434 \pm 0.0002$ whereas the blurred sampler produces $\epsilon_c=0.0338 \pm 0.0003$.
  (c)~Adaptive step size $\Delta s$ versus anneal parameter $s$ for cached and uncached RK45 on the biclique $2\times 6\times 6$ instance. Inset shows cumulative wall time for the simulation, which is greatly reduced with the cached RK45 integrator.
  }
  \label{fig:combined}
\end{figure*}
To suppress this failure mode, we use a single set of Monte Carlo samples across all stages of a trial step.  At the first stage, with parameters $\btheta$, we draw blurred configurations $\{\by_n\}_{n=1}^{N_s}$, store their blur weights $w_n$ (see Eq.~\eqref{eq:blur-reweight}), and cache the log amplitudes $\log \psi_{\btheta}(\by_n)$. At a later stage $i$ with parameters $\btheta_i$, the same configurations are reused with weights
\begin{align*}
  \widetilde{w}_n(\btheta_i)
  &=
  w_n
  \exp\left\{
    2\Repart\left[
      \log \psi_{\btheta_i}(\by_n)-
      \log \psi_{\btheta}(\by_n)
    \right]
  \right\}.
\end{align*}
This construction strongly correlates the Monte Carlo fluctuations between stages, thereby suppressing their contribution to the embedded error estimate.

\section{Results}

Combining the methodological ingredients introduced in the previous section, we obtain stable, accurate, and computationally efficient t-VMC simulations of the D-Wave annealing protocol across all topologies and across different disorder realizations. We systematically increase the expressivity of the variational ansatz through its rank $R$, which leaves the selected body orders unchanged while enlarging the family of representable coefficient tensors. In Fig.~\ref{fig:final}, we show that increasing $R$ allows for the systematic improvement of the correlation errors and the subsequent successful simulation of the Ising anneal for all topologies, at both $t_a=7$ ns and $t_a=20$ ns. 
Our method struggles to systematically improve the correlation errors for the 3D dimer at larger system sizes. This is in line with tensor network results of Ref.~\cite{Tindall2026TensorNetworks}, which found this topology the most challenging to simulate. However, we note that all results in Fig.~\ref{fig:final} were obtained with the same hyperparameters, so a more fine-tuned choice of the regularization and sampling settings could potentially drive down the errors. We conclude from these experiments that the diamond topology is the easiest for t-VMC simulate, and the observed difficulty increases in the order diamond, 2D cylinder, biclique, and 3D dimer. 

\paragraph{Sampler comparison.}

To assess the quality of the sampler, we can monitor three complementary diagnostics. First, we can track the acceptance rate within each replica. High-temperature replicas must maintain sufficiently high local acceptance rates throughout the anneal to explore the state space effectively. Second, we can calculate the adjacent-replica swap rate to determine whether configurations from high-temperature replicas are being exchanged to low-temperature ones. Finally, we can calculate the split-$\widehat{R}$ statistic, which compares between-chain and within-chain fluctuations for representative scalar observables~\cite{Gelman1992Inference,Vehtari2021RankNormalization}. We illustrate these sampling problems using a biclique instance, for which high connectivity and frustration make poor mixing especially pronounced. In Fig.~\ref{fig:combined}(a), we show the freezing of the local Monte Carlo dynamics, which results in a large final $\epsilon_c$. Parallel tempering maintains better mixing and restores the target correlation accuracy to the $<5\%$ range.

\paragraph{Blurred sampling.}

Near the crossover into the glassy regime, the probability mass on spin-flip-connected configurations becomes highly uneven. Standard sampling then produces a sharp deterioration in sample quality and can drive the variational trajectory onto the wrong dynamics. Blurring extends the 
\begin{figure*}[htb!]
  \centering
  \includegraphics[width=\linewidth]{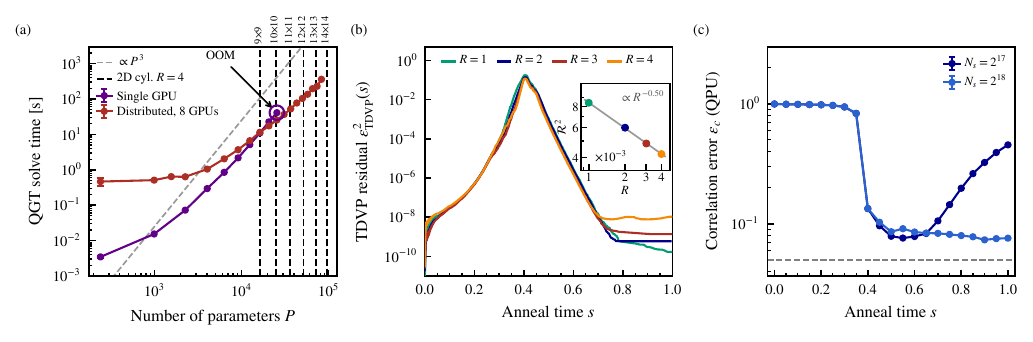}
  \caption{Scaling analysis of the t-VMC simulation (a) Inverting the QGT of the correlator state with $\mcK=(2,4)$ and $R=1$ for system sizes $N = \{16,32,\ldots, 320\}$. Error bars show the standard deviation over 10 random initializations. The vertical dashed line indicates the number of parameters at which a single NVIDIA H200 would run out of memory. (b) Squared TDVP residual $\varepsilon^2_{\mathrm{TDVP}}(s)$ along the anneal for the $8\times8$ 2D cylinder. Curves are smoothed using a rolling window of 100 steps. The inset shows the integrated TDVP error $\mathcal{R}^2= \int_0^1 \varepsilon_{\mathrm{TDVP}}^2(s)\dd s$ as a function of rank; the error decreases approximately as $R^{-0.50}$. (c) Correlation error compared to the QPU as a function of anneal time for a $2\times18\times 18$ biclique instance at $20$ ns. We use a rank $R=2$ correlator ansatz which is sufficient to drive down the errors when the number of samples taken is large enough. Final errors are estimated with $2^{22}=4194304$ samples.}
  \label{fig:scaling}
\end{figure*}

sampled support to Hamiltonian-connected configurations and stabilizes the reweighted estimators. Fig.~\ref{fig:combined}(b) illustrates this behavior through the deterioration of the split-$\widehat{R}$ diagnostic for the energy estimator under standard sampling.

\paragraph{Cached RK45 integrator.}

We isolate the effect of caching the integration stages on the biclique instance by toggling the cache while holding all other hyperparameters fixed. As shown in Fig.~\ref{fig:combined}(c), common-sample reweighting stabilizes the error estimate, permits larger accepted steps, and reduces the number of expensive Markov-chain updates per unit anneal time. This results in a scalable scheme across large system sizes (see App.~\ref{app:adaptive}). 


\paragraph{Distributed QGT solve.}

For the largest ans\"atze, the dense QGT becomes the dominant memory object. If $P$ denotes the number of complex parameters used by the linear solver, the matrix requires $\mathcal{O}(P^2)$ storage, and its eigendecomposition requires additional workspace. A single-device dense solve can therefore become the computational bottleneck even when sampling remains affordable. To overcome this limitation, we use JAXMg~\cite{wiersema2026jaxmg} to distribute the QGT eigendecomposition across any number of GPUs and nodes, which removes the single-device memory
ceiling for the QGT solve and extends the accessible parameter range, see Fig.~\ref{fig:scaling}(a). 

\paragraph{The TDVP projection error.}

The TDVP projection error is calculated by comparing the exact Schr\"odinger velocity with the tangent-space velocity generated by the regularized update~\cite{Schmitt2020QuantumDynamics}:
\begin{align}
	\varepsilon_{\mathrm{TDVP}}^2(s)
	=
	1+
	\frac{
		\dot\btheta^\dagger S(\btheta)\dot\btheta
        -2\Impart(\dot\btheta^\dagger\bF(s,\btheta))
	}{
		\Var[H(s)]
	} .
	\label{eq:tdvp-error}
\end{align}
Residuals of this form are commonly used in t-VMC~\cite{Carleo2017Neural,Schmitt2020QuantumDynamics, Hofmann2022,Mauron2025Challenging}, and tensor-network TDVP~\cite{haegeman2016unifying, ho2019periodic,michailidis2020slow, li2024time, cecile2024measurement} as a diagnostic of expressivity error in the variational ansatz. While useful in practice for monitoring the quality of a simulation, it does not include time-integration, sampling, or linear-solver errors so care must be taken to interpret this quantity in the context of t-VMC. 
In Fig.~\ref{fig:scaling}(b), we observe that $\varepsilon_{\mathrm{TDVP}}^2(s)$ peaks around the critical point. Moreover, the integrated TDVP residual can be systematically reduced by increasing the ansatz rank. This behavior is consistent with the interpretation of the residual as a measure of insufficient variational expressivity, which improves the final correlation error $\epsilon_c$. We investigate this connection in more detail in App.~\ref{app:tdvp}.

\paragraph{Large scale examples} To illustrate the scalability of our approach, we study a challenging $N=72$ biclique instance and compare with the data from the QPU~\cite{kingPrivate}. At this scale, we have to double the number of samples compared to the results of Fig.~\ref{fig:final} in order to accurately resolve the forces. We find a $\approx 7.6\%$ error between the t-VMC simulation and the QPU, which is reasonable given the noise of the QPU. Finally, we study an $N=128$ diamond instance in App.~\ref{app:diamond_large} to further demonstrate the scalability and stability of our simulations. Although we do not have ground truth at this scale, the integrated TDVP error of $\mathcal{R}^2=4.27\times 10^{-3}$ indicates a highly stable simulation.

\section{Discussion and outlook}
\label{sec:discussion}

The calculations presented here demonstrate that, for every system size and graph topology studied here, t-VMC can approximate the final two-spin-correlation error reported for the D-Wave QPU. This agreement improves systematically as the expressive power of a correlator-state ansatz is increased. Our ablation studies identify three ingredients required for controlled t-VMC evolution in the regimes examined. First, the variational family must contain sufficiently many higher-order correlator channels to represent the correlations generated during the anneal. Second, the Monte Carlo sampler must remain equilibrated as the system enters the glassy regime: parallel tempering mitigates slow mixing, while blurred sampling reduces high-variance local-energy estimates and support-mismatch instabilities. Third, importance-weighted adaptive integration enables larger stable time steps, substantially reducing the cost of the time evolution. Together, these developments provide a practical foundation for studying more demanding nonequilibrium dynamics with more powerful variational ans\"atze, including NQS. Additionally, our methodological improvements carry over to tensor network variational families, for which t-VMC has recently attracted renewed interest~\cite{SandvikVidal2007TensorVMC,schuch2008simulation,ferris2012variational,Wu2026real,Dziarmaga2026PEPSAnnealing,HryniukSzymanska2026MPOTVMC}.

Our results establish t-VMC as a competitive complement to tensor-network approaches for simulating nonequilibrium transverse-field Ising annealing protocols. 
Notably, our approach accurately treats the biclique instances at scale, which were not considered in the recent belief-propagation study of Ref.~\cite{Tindall2026TensorNetworks} and believed to be challenging with such approaches~\cite{nocera2025evaluating}. At the same time, the present simulations remain computationally demanding. In our current implementation, the dominant cost is the dense solution of the QGT linear system, which scales as $\mathcal{O}(P^3)$. Because the variational ansatz used here contains $P=\mathcal{O}(N^2)$ parameters, this step has a nominal scaling of $\mathcal{O}(N^6)$. JAXMg~\cite{wiersema2026jaxmg} distributes this computation across many GPUs but cannot eliminate the rapid growth in total computational cost and memory. Extrapolating the present implementation to the largest and most challenging instances of Ref.~\cite{King2025BeyondClassical} is therefore still prohibitively expensive. This scaling should, however, be understood as a property of the present ansatz rather than as a fundamental lower bound on t-VMC; further methodological improvements may shift this computational boundary. We also leave open the question of studying the $\pm1$ instances, which are potentially more challenging for t-VMC.

These findings sharpen, rather than settle, the question of quantum advantage. Our calculations show that the correlations of selected two-spin observables can be reproduced classically at an error level of on par with the QPU, albeit at significant computational costs. In particular, our simulations took hundreds of GPU hours, whereas the QPU obtains the same results in seconds. Rather than viewing classical simulation solely as a challenge to quantum advantage, our results illustrate how the continuing exchange between quantum hardware and classical algorithms catalyzes methodological advances on both sides. We hope that this productive interplay will open access to previously unexplored regimes of quantum many-body dynamics and ultimately lead to the discovery of new physics.

\begin{acknowledgments}

I thank Wladislaw Krinitsin and Markus Schmitt for identifying a methodological flaw in an earlier version of the simulations~\cite{Krinitsin2026comment} related to the final correlation errors. Additionally, I am grateful to Filippo Vicentini and Alessandro Santini for suggesting the cached RK45 method.
I am grateful to Juan Carrasquilla, Andrew King and Jack Raymond for the discussions during the completion of this work.
I also want to thank and Joey Tindall for his comments on the final draft of this manuscript.
The Flatiron Institute is a division of the Simons Foundation.
\end{acknowledgments}

\bibliographystyle{unsrt}
\bibliography{library}

\clearpage
\onecolumngrid
\appendix
\section{Topologies\label{app:topologies}}

\begin{table}[htbp]
  \centering
  \caption{Spin-glass topologies and system sizes. All instances use precision~$256$ couplings, which means the couplings $w_{ij}$ have values in $[-1,+1]$ discretized to a step size of $1/256$ (excepting $0$).}
  \label{tab:topologies}
  \begin{tabular}{@{}lccc@{}}
    \toprule
    Shape & $N$ & \#Bonds &  $d_{\max}$ \\
    \midrule
    \multicolumn{4}{@{}l}{\emph{2D cylinder}} \\
    $4\times4$ & $16$ & $28$  &  $4$ \\
    $5\times5$ & $25$ & $45$  &  $4$ \\
    $6\times6$ & $36$ & $66$  &  $4$ \\
    $7\times7$ & $49$ & $91$  & $4$ \\
    $8\times8$ & $64$ & $120$ & $4$ \\
    \midrule
    \multicolumn{4}{@{}l}{\emph{3D dimer}} \\
    $3\times2\times2\qquad$ & $24$ & $38$   & $4$ \\
    $3\times2\times3$ & $36$ & $75$   & $5$ \\
    $3\times3\times3$ & $54$ & $117$  & $5$ \\
    \midrule
    \multicolumn{4}{@{}l}{\emph{Biclique}} \\
    $2\times5\times5$ & $20$ & $35$  & $6$ \\
    $2\times6\times6$ & $24$ & $48$  & $5$ \\
    $2\times7\times7$ & $28$ & $63$  & $6$ \\
    $2\times8\times8$ & $32$ & $80$  & $7$ \\
    $2\times9\times9$ & $36$ & $99$  & $9$ \\
    $2\times18\times18$ & $72$ & $360$  & $14$ \\
    \midrule
    \multicolumn{4}{@{}l}{\emph{Diamond}} \\
    $3\times3\times8$ & $18$ & $24$  & $4$ \\
    $4\times4\times8$ & $32$ & $48$ & $4$ \\
    $8\times8\times8$ & $128$ & $224$ & $4$ \\
    \bottomrule
  \end{tabular}
\end{table}
\clearpage
\section{Experiment hyperparameters\label{app:hyperparameters}}

All experiments were performed on either 4 NVIDIA A100s or 8 NVIDIA H200s, based on availability. The total wall time of all experiments was $2123.02$ hours. The most costly simulation (single instance of 2D cylinder, $R=4$ at $20$ ns) took approximately $109.25$ hours on 8 NVIDIA H200 GPUs. This time does not include experimentation during the initial stages of this work.

\begin{table}[htbp]
  \centering
  \caption{VMC hyperparameters for the $s=0$ ground state preparation.}
  \label{tab:hp-vmc}
  \begin{tabular}{@{}lll@{}}
    \toprule
    Parameter & Value & Meaning \\
    \midrule
    \texttt{n\_vmc\_steps} & $1000$              & SR optimization iterations \\
    \texttt{vmc\_lr}       & $10^{-3}$           & Initial learning rate \\
    optimizer              & SGD                 & \\
    LR schedule            & cosine decay        & Decay to $10^{-4}$ \\
    \texttt{diag\_shift}   & $10^{-5}$           & QGT diagonal shift $S(\btheta) + \lambda I$ \\
    V-score threshold      & $10^{-4}$           & Convergence criterion \\
    \bottomrule
  \end{tabular}
\end{table}

\begin{table}[htbp]
  \centering
  \caption{Common settings shared by every experiment unless stated otherwise.}
  \label{tab:hp-common}
  \begin{tabular}{@{}lll@{}}
    \toprule
    Parameter & Value & Meaning \\
    \midrule
    \multicolumn{3}{@{}l}{\emph{Ansatz}} \\
    \texttt{orders}               & $[2,4]$                                       & Correlator body orders $k$\\
    \texttt{param\_dtype}         & complex                                       & Holomorphic parameters \\
    \texttt{param\_initializer}   & $\mathcal{N}\!\big(0,(0.01/N)^2\big)$          & Same across all parameters \\
    \midrule
    \multicolumn{3}{@{}l}{\emph{Monte Carlo sampling}} \\
    \texttt{n\_samples}     & $2^{17}=131072$    & MCMC samples \\
    \texttt{n\_chains}      & $2048$             & Markov chains \\
    \texttt{sampler}        & \texttt{pt}        & Parallel tempering with local spin flips\\
    \texttt{pt\_n\_replicas}& $16$               & Replicas per chain along the $\beta$-ladder \\
    \texttt{sweep\_size}    & $4\,N$ & MC sweeps between samples \\
    $\beta_0$     & 1 & Inverse temperature of physical system \\
    $\beta_{N_T-1}$   & 1/16 & Inverse temperature of hottest system replica \\
    \texttt{beta\_spacing}    & \texttt{linear} & $\beta_k$ values are spaced linearly between \texttt{beta\_min} and \texttt{beta\_max}\\
    $q$             & $0.3$      & See~\cite{Wan2026BlurredSampling} for details. Simulations are robust to the choice of $q$\\
    \midrule
    \multicolumn{3}{@{}l}{\emph{TDVP solve}} \\
    \texttt{rcond}         & $10^{-14}$ & \\
    \texttt{rcond\_smooth} & $10^{-12}$ & See~\cite{medvidovic2023variational} for details.\\
    \texttt{atol}          & $1$        & See~\cite{Schmitt2020QuantumDynamics} for details.\\
    \midrule
    \multicolumn{3}{@{}l}{\emph{Integrator}} \\
    \texttt{integrator} & \texttt{rk45} & Adaptive Dormand--Prince RK45 method~\cite{Dormand1980Family} with \texttt{rtol}$=10^{-5}$ \\
    \texttt{dt\_min}    & $10^{-4}$     & Minimal step size \\
    \texttt{dt\_max}    & $10^{-2}$     & Maximal step size \\
    \texttt{cache\_within\_step} & \texttt{True} & Reuse samples across RK stages \\
    \bottomrule
  \end{tabular}
\end{table}

\begin{table}[htbp]
  \centering
  \begin{threeparttable}
  \caption{t-VMC hyperparameters for individual figures.}
  \label{tab:hp-fig1}
  \begin{tabular}{@{}lll@{}}
    \toprule
    Parameter & Value  \\
    \midrule
    \multicolumn{3}{@{}l}{ Fig.~\ref{fig:final}} \\
    \texttt{topology}  & \{\texttt{2d}, \texttt{diamond}, \texttt{3ddimer}, \texttt{biclique}\} & \\
      $R$    & $\{1,2,3,4\}$                    &  \\
    $t_a$      & $\{7,20\}$ ns                    & \\
    \midrule
    \multicolumn{3}{@{}l}{ Fig.~\ref{fig:combined}(a)} \\
    \texttt{topology} & \texttt{biclique} & \\
    \texttt{shape}    & $2\times6\times6$         & \\
     $R$   & $3$               & \\
    $t_a$     & $7$ ns            & \\
    \texttt{sampler}  & \{\texttt{local}, \texttt{pt}\} & \\
    \midrule
    \multicolumn{3}{@{}l}{ Fig.~\ref{fig:combined}(b)} \\
    \texttt{topology} & \texttt{biclique} & \\
    \texttt{shape}    & $2\times6\times6$         & \\
     $R$     & $3$               & \\
    $t_a$     & $7$ ns            & \\
    $q$        & $\{0.0,\,0.3\}$   & \\
    \midrule
    \multicolumn{3}{@{}l}{ Fig.~\ref{fig:combined}(c)} \\
    \texttt{topology} & \texttt{biclique} & \\
    \texttt{shape}    & $2\times6\times6$         & \\
     $R$   & $3$               & \\
    $t_a$     & $7$ ns            & \\
    \texttt{cache\_within\_step} & \{\texttt{True}, \texttt{False}\} &  \\
    \midrule
    \multicolumn{3}{@{}l}{ Fig.~\ref{fig:scaling}(b)} \\
    \texttt{topology} & \texttt{2d} & \\
    \texttt{shape}    & $8\times 8$         & \\
     $R$   & $\{1,2,3,4\}$               & \\
    $t_a$     & $20$ ns            &\\
    \midrule
    \multicolumn{3}{@{}l}{ Fig.~\ref{fig:scaling}(c)} \\
    \texttt{topology} & \texttt{biclique} & \\
    \texttt{shape}    & $2\times 18\times 18$         & \\
     $R$   & $2$               & \\
    $t_a$     & $20$ ns            &\\
    \texttt{n\_samples}     & $\{2^{17}, 2^{18}\}$           &\\
    \midrule
    \multicolumn{3}{@{}l}{ Fig.~\ref{fig:large_diamond}}\\
    \texttt{topology} & \texttt{diamond} & \\
    \texttt{shape}    & $8\times 8\times 8$         & \\
     $R$   & $1$               & \\
    $t_a$     & $7$ ns            &\\
    \bottomrule
  \end{tabular}
  \end{threeparttable}
\end{table}

\clearpage
\section{Miscellaneous numerical details\label{app:failed}}

\subsection{Additional ans\"atze}
We experimented with several extensions to the correlator state ansatz of~\Cref{sec:ansatz}. 
Adding a cRBM to the ansatz produced a TDVP equation that was substantially more difficult to regularize, requiring a large $\texttt{rcond\_smooth}\approx10^{-6}$ to stabilize the dynamics and exhibiting strong sensitivity to the regularization~\cite{Hofmann2022}. Nonetheless, the additional expressivity was insufficient to capture the correct correlation structure for most instances. Adding terms that explicitly parameterize correlators associated with $m$-cycles in the problem graph—for example, all 4-cycles on the diamond lattice—also did not improve the final results.

\subsection{Regularization}
Regularizing the TDVP equation is at the heart of t-VMC and can greatly affect the quality of the numerical results. This is especially problematic for NQS and other highly nonlinear ans\"atze (unlike in tensor networks, where such regularization can be done in a controlled way~\cite{Wu2026real}).
The correlator state is quite robust to regularization across all topologies. As long as \texttt{rcond\_smooth} was chosen to be less than $\approx 10^{-10}$, the dynamics were stable and produced correlation errors on par with the final hyperparameter used in the main text.

\subsection{Mixed precision}
In~\cite{Solinas2026neural}, we observed that mixed precision computing can be employed to reduce optimization costs in VMC ground state searches. In particular, sampling the variational state can be done in low-precision data formats (like \texttt{f32} or even \texttt{f16}), without meaningfully affecting the trajectories. 
The reason is that VMC calculations typically use a small number of samples, such that the statistical noise in the force and QGT estimators can exceed the bias introduced by reduced sampling precision.
In t-VMC, we found that this is not the case. We require extremely precise estimates in order to resolve the dynamics accurately at every step. Numerical experiments with mixed-precision sampling for the results in the main text indicated significant sampling speedups, but resulted in numerical errors that affected the stability of the adaptive RK solver and ultimately the resulting dynamics.

\subsection{ODE integration failures}
The TDVP equation can be extremely stiff~\cite{vrcan2025instability}, hence on rare occasions the ODE integration can fail. We only observed this for the $20$ ns biclique instances, where sampling and support mismatch are especially pathological. When this happens, we revert the state of the ODE solver, thermalize the Markov chains for $5\times$ the current sweep size, and try again, which was enough to recover the correct dynamics. Additional ODE integration errors can occur due to spurious samples destroying the force estimate. Similarly, we only experienced these issues in the final production runs of the $20$ ns biclique instances. Ideally, we would remove extreme outliers in the force estimators at the cost of incurring a small bias. However, since we did not want to repeat the entire simulation with this change, we instead restarted the crashed simulation whenever this rare event occurred, which resolved the issue in practice.
\clearpage

\section{An AD-friendly parameterization of a strictly upper-triangular matrix}
\label{app:filltri}

In this section, we describe a numerical trick for efficiently parameterizing an upper-triangular matrix that, to our knowledge, has not previously been used in this context.

Each channel matrix $U^{(k,r)}\in\mathbb{C}^{N\times N}$ in Eq.~\eqref{eq:path-factor} is strictly upper triangular and therefore contains $m=\frac{N(N-1)}{2}$
free complex entries. Storing a dense $N\times N$ matrix and masking its lower triangle introduces unused parameters, which can introduce additional sources of MCMC noise. Conversely, scattering a flat parameter vector into a zero matrix can be inefficient for the automatic-differentiation machinery used in our implementation. 

We can instead construct the upper-triangular matrix with dense concatenation, reshape, and masking operations. Let $\bv\in\mathbb{C}^{m}$ contain the free parameters and define the concatenated vector
\begin{equation*}
  \widetilde{\bv}
  = \left[\bv,\rev(\bv),\bzero_N\right]\in\mathbb{C}^{N^2}.
\end{equation*}
Using row-major ordering, the desired matrix is
\begin{equation}
  U(\bv)
  = \triu\!\left(
    \operatorname{reshape}(\widetilde{\bv},(N,N)),1
  \right),
  \label{eq:filltri}
\end{equation}
where the second argument indicates that only entries strictly above the diagonal are retained. This construction is a permutation of the entries of $\bv$ within the strict upper triangle, so every parameter appears exactly once.

We illustrate this idea with an $N=4$ example. We have the vectors
\begin{align*}
  \bv&=\left(v_1, v_2, v_3, v_4, v_5, v_6\right),\\
  \widetilde{\bv}
  &=
  \left(
  v_1,v_2,v_3,v_4,v_5,v_6,
  v_6,v_5,v_4,v_3,v_2,v_1,
  0,0,0,0
  \right).
\end{align*}
A row-major reshape gives
\begin{equation*}
  \operatorname{reshape}(\widetilde{\bv},(4,4))
  =
  \begin{pmatrix}
    v_1 & v_2 & v_3 & v_4 \\
    v_5 & v_6 & v_6 & v_5 \\
    v_4 & v_3 & v_2 & v_1 \\
    0   & 0   & 0   & 0
  \end{pmatrix},
\end{equation*}
and retaining the strict upper triangle yields
\begin{equation*}
  U(\bv)
  =
  \begin{pmatrix}
    0 & v_2 & v_3 & v_4 \\
    0 & 0   & v_6 & v_5 \\
    0 & 0   & 0   & v_1 \\
    0 & 0   & 0   & 0
  \end{pmatrix}.
\end{equation*}
Thus the operation in Eq.~\eqref{eq:filltri} avoids indexed scatter operations while preserving a one-to-one parameterization of the strict upper triangle.

\clearpage
\section{Scaling of the residual TDVP error \label{app:tdvp}}

\begin{figure*}[htb!]
    \centering
    \includegraphics[width=\linewidth]{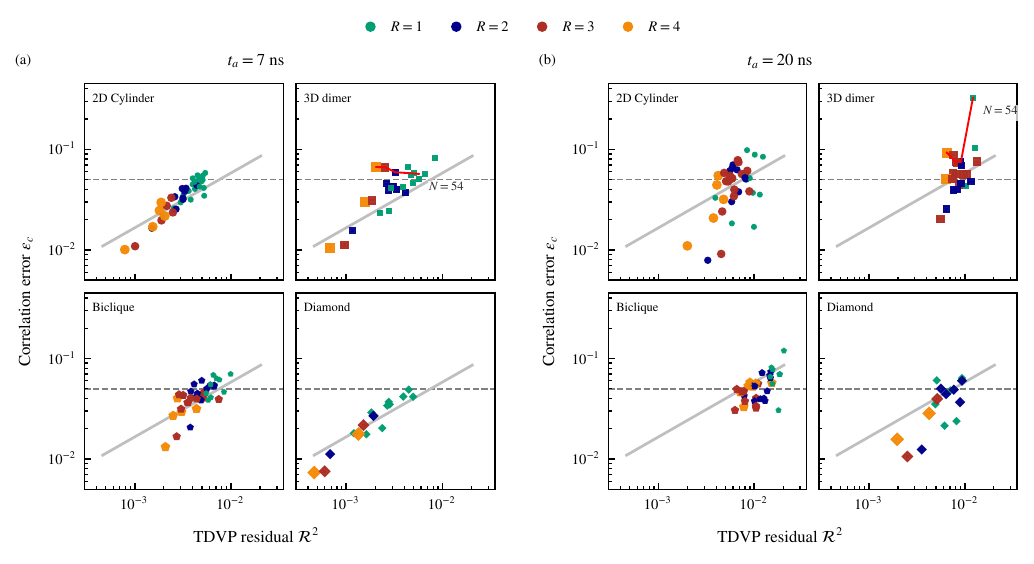}
    \caption{
    Predicting the correlation error from the integrated TDVP error. The markers are colored by ansatz rank $R\in\{1,2,3,4\}$. The integrated TDVP error $\mathcal{R}^2$ is scattered against the final-time correlation error $\epsilon_c$ for annealing times (a) $7$ ns and (b) $20$ ns across all system sizes. The dashed line is the $5\%$ target, and the gray solid line indicates a log-log linear relation to guide the eye. 
    }
    \label{fig:tdvp_error_epsc}
\end{figure*}

It has been proposed that accuracy of t-VMC simulations can be derived from the value of the integrated TDVP error:
\begin{align}
    \mathcal{R}^2= \int_0^1 \varepsilon_{\mathrm{TDVP}}^2(s)\dd s
\end{align}
In Fig.~\ref{fig:tdvp_error_epsc} we show how the final correlation error changes as a function of the $\mathcal{R}^2$ over all experiments. 
For $7$ ns, there seems to be a strong correlation between the residual TDVP error and final accuracy of the simulation. In particular, for the diamond instances at $7$ ns the correlation error seems to scale linearly with $\mathcal{R}^2$, which is in line with the observations of Ref.~\cite{Mauron2025Challenging}. However, for the $20$ ns quenches, the correlation between the TDVP error and $\epsilon_c$ is much less clear.
Furthermore, for the 3D dimer instances of size $N=54$, we observe that $\mathcal{R}^2$ decreases while $\epsilon_c$ increases, providing a counterexample to the hypothesis that $\mathcal{R}^2$ controls the simulation accuracy. Stronger regularization, more samples or increased sweep sizes could potentially correct this behavior, but brief experiments proved incapable of recovering the trend seen for the other topologies. 
From our experiments, it is therefore unclear how strong of an indicator the TDVP error is for the success of a simulation and whether it can be used to extrapolate the results across all topologies and timescales.

\begin{figure*}[htb!]
    \centering
    \includegraphics[width=\linewidth]{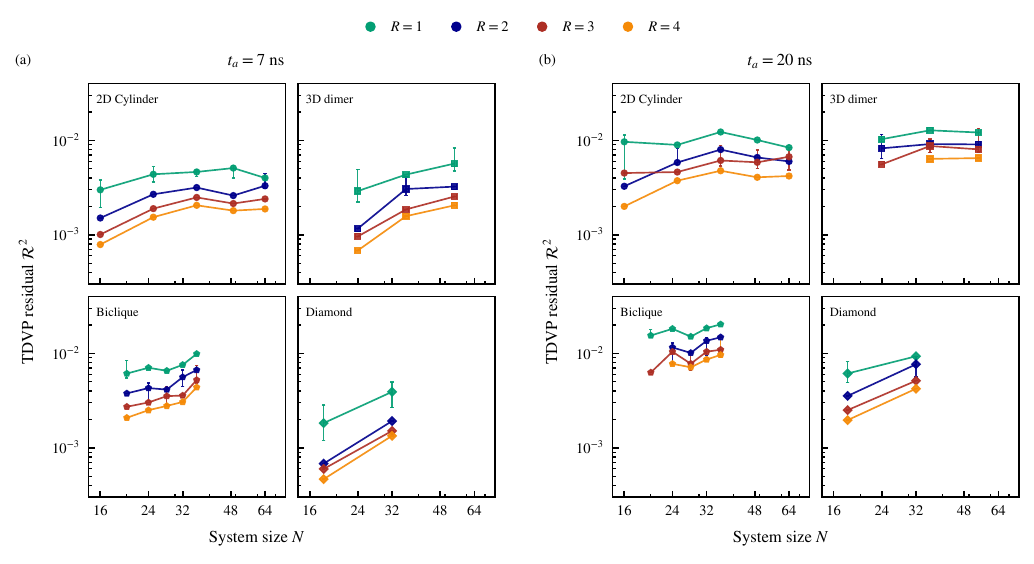}
    \caption{
    The integrated TDVP error $\mathcal{R}^2$ as a function of system size for both
    (a) $7$ ns and (b) $20$ ns.
    Some data are missing for the 3D dimer and biclique quenches, where spurious samples prevented the accurate estimation of $\varepsilon_{\mathrm{TDVP}}^2(s)$.  The ordering of the rank lines shows the systematic improvement of  $\mathcal{R}^2$ with the ansatz rank $R$. The error bars indicate the minimum and maximum values, with the median over 5 disorder instances indicated by the point itself.
    }
    \label{fig:tdvp_error_size}
\end{figure*}

In Fig.~\ref{fig:tdvp_error_size}, we investigate how the integrated TDVP error scales as a function of system size. While the value of $\mathcal{R}^2$ seems to flatten for the 2D topology, it is unclear what will happen for biclique and diamond lattices. 

From these data, we believe that it is not possible to make strong statements about the scalability of these simulations. It is likely that larger system sizes require more samples to resolve the TDVP equation accurately and that sweep sizes have to be increased to decorrelate Monte Carlo chains in the glassy regime. Identifying the scaling of these hyperparameters is crucial to determine the scalability of our simulations, but is left to future work.

\clearpage
\section{Scaling of adaptive time step\label{app:adaptive}}
\begin{figure*}[htb!]
    \centering
    \includegraphics[width=\linewidth]{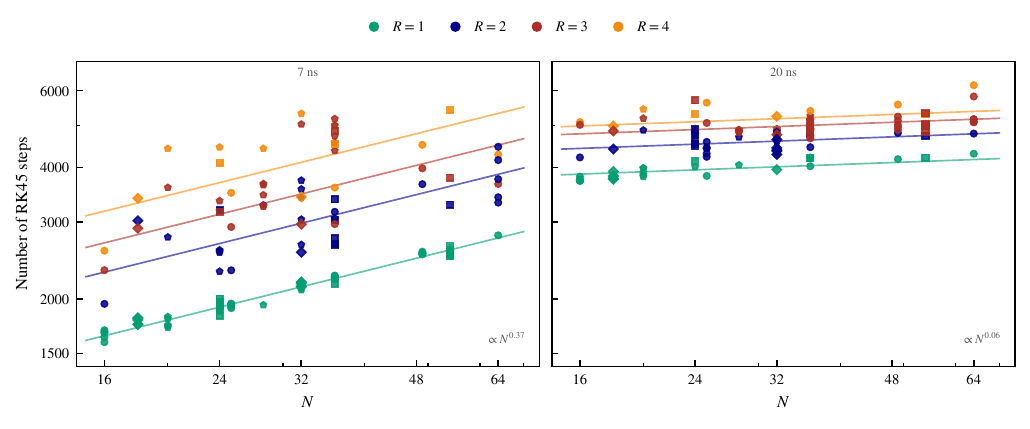}
    \caption{Scaling of the number of Runge--Kutta integration steps as a function of system size. For each rank, the data follow a linear trend, with a constant offset as a result of increasing rank of the ansatz.
    }
    \label{fig:step size}
\end{figure*}

In this section, we investigate how the number of required ODE integration steps scales with system size. In Fig.~\ref{fig:step size} we find that the number of steps scales as $N^{0.37}$ for $7$ ns, whereas for $20$ ns, the number of steps is roughly constant.

\clearpage

\section{Detailed information of large-scale diamond instance\label{app:diamond_large}}

\begin{figure}[htb!]
    \centering
    \includegraphics[width=\linewidth]{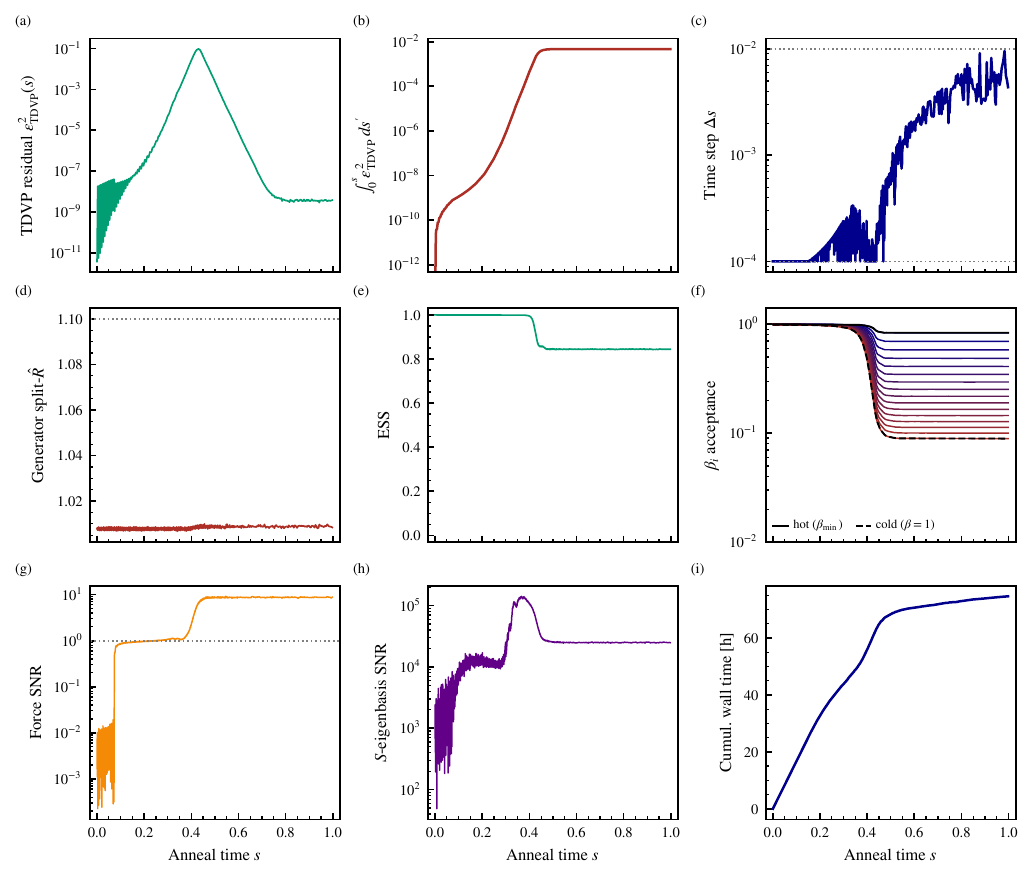}
    \caption{Run diagnostics for a t-VMC simulation of the $N=128$ diamond instance at $t_a=7$~ns with a rank-$R{=}1$ ansatz.
    (a) TDVP residual $\epsilon^2_{\mathrm{TDVP}}(s)$.
    (b) The integrated TDVP error $\mathcal{R}^2$ over time, which reaches a final error of $\mathcal{R}^2=4.27\times 10^{-3}$.
    (c) The adaptive step $\Delta s$ between the bounds $ds_{\min}=10^{-4}$ and
    $ds_{\max}=10^{-2}$ (dotted).
    (d) The split-$\hat{R}$ diagonistic of the TDVP generator compared with the $\hat{R}<1.1$
    threshold (dotted);
    (e) The effective sample size of the blurred estimator.
    (f) The local acceptance of all $16$ tempering replicas, hot (solid black) to cold
    (dashed black).
    (g) The median signal-to-noise ratio of the force in the parameter basis against the filter
    floor $\mathrm{snr}_{\mathrm{atol}}=1$ (dotted), and (h) the median signal-to-noise in the
    eigenbasis of the quantum geometric tensor $S$.
    (i) The cumulative wall time.}
    \label{fig:large_diamond}
\end{figure}

\end{document}